**Noise-immune and AI-enhanced DNA storage via adaptive partition mapping of digital data**

Zimu Li[1], Bingyi Liu[1], Lei Zhao[1,*], Qian Zhang[1], Yang Liu[1], Jun Liu[3], Ke Ke[3], Huating Kong[4], Xiaolei Zuo[2], Chunhai Fan[1,2], Fei Wang[1,*]

[1]State Key Laboratory of Synergistic Chem-Bio Synthesis, School of Chemistry and Chemical Engineering, Frontiers Science Center for Transformative Molecules, Institute of Translational Medicine, Shanghai Jiao Tong University, Shanghai 200240, China
[2]Institute of Molecular Medicine, Shanghai Key Laboratory for Nucleic Acids Chemistry and Nanomedicine, Renji Hospital, School of Medicine, Shanghai Jiao Tong University, Shanghai 200127, China
[3]Lenovo Research, Lenovo Group, Beijing 100094, China
[4]Shanghai Synchrotron Radiation Facility (SSRF), Shanghai Advanced Research Institute, Chinese Academy of Sciences, Shanghai, 201204, China

*Correspondence: l.zhao@sjtu.edu.cn ; wangfeu@sjtu.edu.cn

**Abstract**
The information age and the rise of artificial intelligence are intensifying demands for massive, durable data storage. Encoding digital information into DNA sequences offers an attractive potential solution for storing rapidly growing data, but practical implementations of DNA storage are constrained by errors introduced during synthesis, preservation, and sequencing processes. To overcome these channel noises, current encoding architectures primarily rely on error-correcting codes to detect and rectify errors However, these systems remain vulnerable to noise levels that exceed predefined thresholds. Here, we developed a Partitioning-mapping with Jump-rotating (PJ) encoding scheme, which exhibits exceptional noise resilience. PJ removes cross-strand information dependencies so that strand loss manifests as localized gaps rather than catastrophic file failure. It prioritizes file decodability under arbitrary noise conditions and leverages AI-based inference to enable controllable recovery of digital information. For the intra-strand encoding, we develop a jump–rotating strategy that relaxes sequence constraints relative to conventional rotating codes and provides tunable information density via an adjustable jump length. Based on this encoding architecture, the original file information can always be decoded and recovered under any strand loss ratio, with fidelity degrading smoothly as damage increases. We demonstrate that original files can be effectively recovered even with 10% strand loss, and machine learning datasets stored under these conditions retain their classification performance. Experiments further confirmed that PJ successfully decodes image files after extreme environmental disturbance using accelerated aging and high-intensity X-ray irradiation. By eliminating reliance on prior error probabilities, PJ establishes a general framework for robust, archival DNA storage capable of withstanding the rigorous conditions of real-world preservation.

**Introduction**
With the rapid advancement of artificial intelligence (AI) technology, we have entered a data-driven era. AI technology generates massive volumes of data and relies on these extensive datasets for learning and evolution[1,2]. The exponential growth of data volume necessitates storage systems with significantly higher capacity. DNA, the primary carrier of genetic information in nature, exhibits exceptionally high information density and stability[3,4]. Progress in DNA synthesis and sequencing technologies now enable the writing of digital data into nucleotide sequences and the subsequent retrieval of the original information[5-7]. Therefore, DNA storage presents an attractive solution to the digital information storage challenges in the context of AI and big data[8-10]. Prior research has successfully utilized DNA to store a diverse array of information formats, including text, images, audio, video, web pages, and even functional computer programs[11-15].

As a biological macromolecule, DNA is inherently prone to errors. Undesirable reactions during synthesis and sequencing, along with degradation during storage, can introduce nucleotide substitutions, insertions, deletions[16,17], and even complete strand dropout[18]. These errors present significant challenges to the reliable recovery of information stored in DNA. To address the challenges, a common approach involves introducing sufficient logical or physical redundancy to correct errors induced by these biochemical reactions[19,20]. Current DNA storage encoding strategies typically map files in their entirety onto a DNA sequence library, where the information carried by different DNA strands is often interdependent[12,21]. These encoding architectures must incorporate a suitable amount of error-correction redundancy which is matched to the noise levels of the storage process. Once the actual noise exceeds the anticipated thresholds, the information stored in DNA becomes irretrievable. However, the channel noise level is highly unpredictable and depends on the specific physical implementation of the storage architecture. In DNA storage workflows, the types and frequencies of errors are significantly influenced by choosing organic or enzyme-mediated synthesis, liquid or powder-based storage formats, and next-generation sequencing (NGS) or nanopore-based sequencing technologies[22-26], thereby limiting the practical application of DNA-based storage. Furthermore, the information density of each encoding architecture must also be taken into consideration. If the actual error rate falls below the predefined redundancy threshold, it leads to an inefficient use of the encoding capacity. Conversely, if the actual error rate exceeds this threshold, it results in decoding failure.

With the rapid improvement of modern artificial intelligence in recent years, current AI models have attained considerable reasoning capabilities, enabling them to comprehend the information conveyed by data even when it is partially incomplete or corrupted. Based on the unique characteristics of DNA storage channels and the data usage requirements in the context of AI, we have designed a universal two-level encoding architecture. At the top level, a Partition Mapping (PM) scheme converts the digital file into a library of DNA strands. The file to be stored is partitioned into multiple independent content blocks, each of which is individually mapped to DNA sequences. This approach achieves partition mapping from the file to DNA storage, wherein the decoding failure of one content block does not affect the retrieval of others. At the bottom level, we developed the Jumping Rotating (JR) rule to encode binary data into DNA sequences. By adjusting the jump step length, this rule controls the maximum permitted length of homopolymers, thereby achieving adaptive information density and compatibility with diverse noise levels in DNA storage workflows. Based on the Partition-mapping with Jump-rotating (PJ) architecture, the entire DNA storage system operates independently of any prior error probabilities and exhibits unrestricted resilience to noise. Leveraging the associative and reasoning capabilities of AI, this framework enables a universal approach to DNA-based data storage.

# Results

**Partition mapping scheme for noise immune decoding**

The Partitioning-mapping with Jump-rotating (PJ) scheme incorporates a two-level encoding architecture, operating at both the DNA strand level and the internal sequence level (Figure 1a). The file to be stored is structured into a matrix composed of independent elements, where the information of each unit is encoded within a single DNA strand. This approach achieves partition mapping from the original file to a DNA-based matrix format. At the sequence level, we introduce a Jump Rotating rule where information density can be adaptively tuned by adjusting the jump step size. Through this two-level design, the architecture balances information density and fidelity while guaranteeing unconditional decodability. Furthermore, missing elements in the file can be recovered through AI-based inference.

Conventional digital-to-DNA mapping strategies often process a file as an indivisible unit (entire mapping). Consequently, damage to even a subset of DNA strands can lead to a complete failure of file decoding (Figure S1). In contrast, within the data flow of the partition mapping scheme (Figure 1b), the file is first structured into a set of independent binary string matrices. Each matrix unit is then converted into a DNA sequence, enabling each DNA strand to independently carry a segment of the information. If any DNA strand is damaged, it does not affect the decoding of other strands, thereby ensuring the recoverability of the file. Taking an image file as an example (Figure 1c), the encoding process begins by decomposing the image into its corresponding pixel matrix. Subsequently, the binary strings from the pixel matrix are encoded into corresponding DNA sequences. To locate the position of each strand within the image, row and column indices are incorporated into the DNA strands alongside the pixel data. During decoding, the sequenced information is sorted according to the row and column indices. In the event of strand loss, zero-padding at the absent locations results in a matrix containing missing values, producing a reconstructed image with visible noise. Furthermore, the noisy image can be restored using an inpainting model. Through this encoding strategy, if partial DNA strand loss occurs during storage, the missing data is locally manifested at its original position in the recovered image, without affecting the retrieval of non-missing pixel information. This approach maximizes the likelihood of successful image reconstruction. The use of an image matrix eliminates the dependency on fragile header structures (Figure S2), thereby substantially improving the robustness of DNA-based storage for image data. Since AI models inherently tolerate a certain degree of data loss in image recognition, the priority on decodability in the Partitioning-Mapping (PM) scheme makes DNA-based storage more compatible with AI-driven workflows.

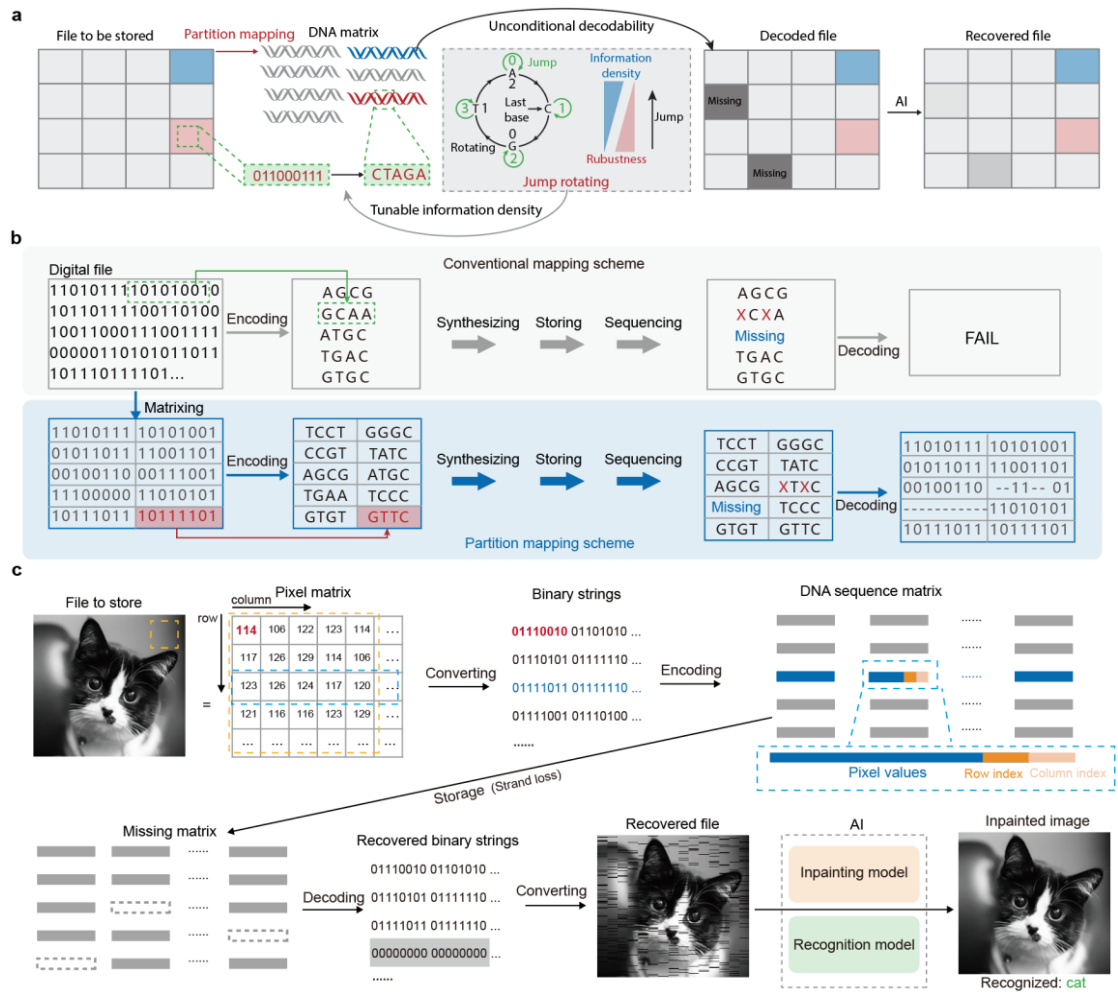

Figure 1 Schematic of noise-immune partition mapping. a, Operational principle of the PJ encoding scheme. b, Data flow of the DNA storage process using the partition mapping encoding strategy. c, Noise-immune DNA storage for image files based on partition mapping. Strand loss does not prevent successful decoding and the reconstructed image is further inpainted by AI.

**Jump rotating rule with adaptive information density**

Following the establishment of the DNA sequence matrix architecture, we subsequently focused on developing encoding rules that map each matrix element to the sequence of a single DNA strand. Here, we propose a jump rotating rule with adaptive information density (Figure 2a and Figure S3). Based on this rule, the selection of the next nucleotide is determined by a rotating rule when the next digital base to be encoded is 3, whereas a direct rule is applied when the next digital base is 4.

In theory, 2-bit encompasses four states and can be directly encoded into a single nucleotide (direct rule), thereby achieving the theoretical maximum encoding density (Figure 2b, direct rule). However, as a biological macromolecule, DNA is subject to specific biochemical constraints. Consequently, not all DNA sequences are permissible for data storage[27]. When a sequence contains long homopolymer regions (e.g., "AAAAA", "CCCCCCC"), the DNA strand becomes unstable, prone to forming secondary structures, and more susceptible to errors during synthesis and sequencing. The rotating encoding scheme, proposed by Goldman et al.[20], encodes a single ternary digit into one nucleotide (Figure 2b, rotating rule). This approach ensures that no homopolymers can form, as it requires that the current nucleotide must differ from the preceding one. This method effectively addresses the constraints of homopolymers yet comes at the cost of reduced encoding density.

Therefore, we developed a hybrid-base method called Jump-rotating (JR) encoding that combines the high encoding density of the quaternary system (direct rule) with the homopolymer avoidance capability of the ternary system (rotating rule). In the JR encoding scheme, a binary string is first converted into a mixed base-3 and base-4 string. It is then mapped to a corresponding DNA sequence through either direct mapping (for quaternary digits) or rotating mapping (for ternary digits), as illustrated in Figure 2c. Compared to the rotating method, the direct mapping of quaternary digits in this approach can be viewed as a "jump" out of the rotating pattern. The jump step size can be represented by the number of quaternary digits located between two ternary digits (Figure 2d). For instance, in $3_4 0_3 1_4 2_3 0_4$, there is one quaternary digit located between the ternary digits '0' and '2', resulting in a 1-jump encoding; in $3_4 0_3 1_4 2_4 0_3$, two quaternary digits are situated between two ternary digits, representing a 2-jump encoding scheme. Figure 2d presents several patterns of DNA sequences under 0, 1, and 2-jump. For example, with a jump step of 1, the first nucleotide is encoded using the rotating rule, while the second is encoded without this constraint—allowing it to be identical to the first. The rotating rule is then reapplied to encode the third nucleotide, ensuring it differs from the second. By iteratively applying the single-step jump-rotating encoding, the resulting sequence is guaranteed to contain at most two consecutive identical nucleotides, thereby entirely preventing the occurrence of homopolymers with three or more repeats. Similarly, increasing the jump step to two ensures that the resulting sequence contains at most three consecutive identical nucleotides. Therefore, an n-jump encoding scheme permits a maximum homopolymer length of n+1. A longer jump step allows for more quaternary digits to be encoded, which correspondingly increases the information density. However, this also introduces a higher risk of errors during DNA synthesis and sequencing. Therefore, this hybrid-base method enables adaptive adjustment of the jump length based on the actual accuracy of the DNA synthesis and sequencing platforms, thereby achieving an optimal balance between encoding density and storage accuracy.

In our encoding strategy for the DNA sequence matrix, we employed a 2-jump encoding scheme, as it accommodates the general tolerance of synthesis and sequencing platforms for up to three consecutive identical nucleotides (homopolymers). In this design, every five nucleotides form a group: the 2nd and 5th positions are encoded in ternary using the rotating rule, while the 1st, 3rd, and 4th positions are encoded in quaternary via direct mapping. This structure enables each 5-nucleotide segment to represent $4 \times 3 \times 4 \times 4 \times 3 = 576$ unique patterns, corresponding to a capacity of 9 bits. Therefore, the binary string is partitioned into 9-bit segments. Each segment is then converted into a 5-digit mixed-base (Figure 2e, insert). Subsequently, these mixed-base digits are transformed into the corresponding nucleotide sequence according to the Jump-Rotating (JR) encoding rules, completing the conversion from binary to DNA. This design achieves a high logical density of 9 bits per 5 nucleotides (1.8 bits/nt) and simultaneously ensures compliance with the homopolymer constraints of DNA sequences. Each synthesized DNA strand carries a 90-nucleotide data region, which is formed by concatenating 18 segments of 5 nucleotides and encodes 162 bits of information. Additionally, row and column indices comprise 10 nucleotides. To facilitate reading with next-generation sequencing (NGS), each DNA strand is flanked by sequencing primers of 20 nucleotides at the 5' end and 21 nucleotides at the 3' end[28] (Figure 2e).

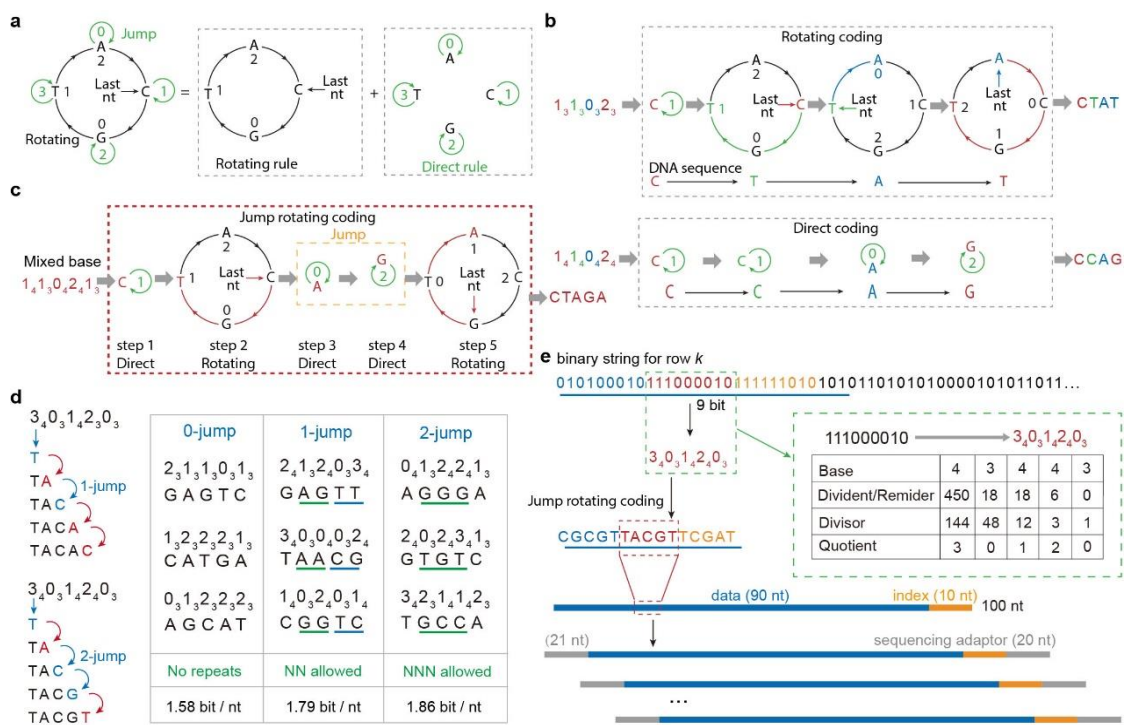

Figure 2 Hybrid-base jump-rotating encoding scheme. (a) Schematic of jump-rotating rule combining the rotating rule and the direct rule. (b) Schematic of ternary rotating coding and quaternary direct coding. (c) Example for converting a mixed-base digit into a DNA sequence via jump-rotating encoding. (d) Encoding process under different jump steps: 0-jump, 1-jump and 2-jump. As jump steps increase, the permitted maximum homopolymer length increases, and the logical encoding density also rises. (e) Binary-to-DNA encoding workflow: binary strings are split into 9-bit segments, each converted via base conversion (green box in Fig. e) to a 5-digit mixed-base number, which is then encoded into a nucleotide sequence. A DNA strand is assembled by a 90-nt data payload, a 10-nt index and ~20-nt primers at both ends.

**Noise immunity of the PJ scheme**

To evaluate the performance of our algorithm in reconstructing images under noisy environment, we performed computer-based simulations of strand dropout scenarios. We selected a portrait of Claude Shannon, the father of information theory (400 × 533 pixels), as a representative example for encoding[29]. After converting the image into an array of DNA strands, each strand was subjected to random loss at a set probability to simulate scenarios involving molecular degradation during storage or incomplete sequencing coverage. By introducing varying degrees of DNA strand loss and subsequently decoding the resulting DNA library, we successfully reconstructed a series of images, as shown in Figure 3a (and Figure S4). As observed in the images, the level of visual degradation increases with higher rates of strand loss. Nonetheless, our algorithm successfully enables the decoding of the file even at a strand loss rate of 75%, demonstrating its robustness in ensuring recoverability under severe data loss conditions, albeit potentially with imperfect visual fidelity.

To further evaluate the noise immunity of the PJ scheme, we compared the reliability of DNA storage using Partition Mapping (PM) versus Entire Mapping (EM) under identical noise conditions. By calculating the Structural Similarity Index (SSIM) of the recovered image and the original image, we can quantitatively assess the deviation of the restored result from the source, thereby evaluating the performance of the algorithm in image recovery. We compared the SSIM of images restored by both PM and EM algorithms under varying strand loss rates through simulations (Figure 3b). We observed that the entire mapping algorithm failed to recover images due to header corruption even at a very low strand loss rate (~0.1%) (Figure S5). In such cases, where image recovery was impossible, the SSIM was assigned a value of 0. In contrast, the partition mapping algorithm exhibited only a gradual decline in image similarity as the strand loss rate increased. Notably, the SSIM never dropped to zero until complete strand loss occurred, demonstrating the robust capability of the PM method to reliably recover files under data loss scenarios.

These simulations were conducted without any post-processing optimization applied to the recovered images. In practice, transitions between pixels in images are typically smooth. Therefore, image inpainting algorithms can be incorporated into the decoding pipeline to enhance visual quality. Here, we introduce an AI-assisted image inpainting program and apply it to refine the defective images recovered by our algorithm. As demonstrated in Figure 3c and Figure S6, the inpainting process not only yields a more natural visual output but also significantly increases the Structural Similarity Index (SSIM) (Figure 3d).

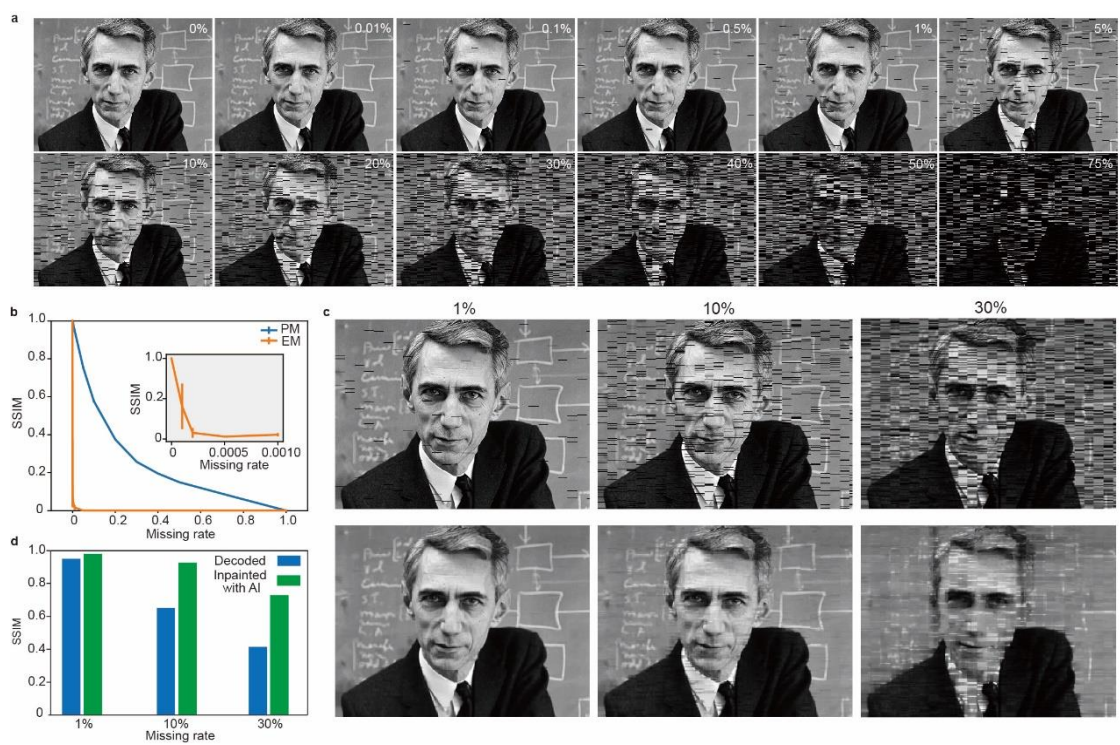

Figure 3 Image recovery performance of PJ encoding scheme under various noise levels. (a) Representative decoded images under different DNA strand loss rates. (b) SSIM between recovered and original images as a function of strand loss rate, comparing partition mapping and conventional entire mapping schemes. (c) Effect of AI inpainting on decoded data under varying loss rates. (d) Comparison of SSIM before and after inpainting for the recovered images shown in panel (c).

**Noise-immune DNA storage for pattern recognition applications**

During image recovery under strand loss, we observed that although the reconstructed images lost a subset of pixels, these defects did not impair the extraction of essential semantic information. For instance, Shannon remained clearly recognizable in the partially damaged portrait, demonstrating that limited image degradation often does not hinder the practical retrieval and application of key visual information in real-world scenarios. A typical application is image recognition. In recent years, with continuous advances in machine learning, the use of image recognition technology has become increasingly widespread. Since image recognition generally focuses on identifying the type (label) of the main subject in an image, minor defects or missing portions often do not affect the model's ability to accurately determine the image label.

To investigate the extent to which image defects affect recognition performance, we developed a machine learning model for image recognition. We selected the Fashion MNIST dataset[30] from the TensorFlow library (Figure 4a). This dataset consists of 70,000 grayscale images across 10 categories of clothing, including 60,000 training images and 10,000 test images, each with a resolution of 28×28 pixels. We first trained the image recognition model using the architecture provided by TensorFlow on the training set, and subsequently evaluated its performance on the test set, achieving a recognition accuracy of approximately 90%. Subsequently, we examined the difference in recognition accuracy of the AI model between intact and partially damaged images. We randomly selected a subset of images from the test set and encoded them using the PJ scheme. The resulting DNA sequences were then subjected to 1% strand loss, after which the recognition accuracy of the degraded images was evaluated (Figure 4b). We observed that the recognition results of AI model for the polluted data remained consistent with those of the original files. We further increased the loss rate to 10% and performed multiple rounds of randomized contamination, each producing images with distinct damage patterns upon decoding (Figure 4c). Although a 10% loss rate resulted in visibly apparent defects in the images, the model consistently produced correct recognition results for every instance. This indicates that, under certain conditions, a moderate level of data loss does not significantly impair the outcome of image recognition tasks.

Subsequently, we encoded all images in the test set using the PJ scheme, introduced a 10% strand loss contamination, and then decoded the data to generate the corresponding set of degraded images. We used the same AI model to perform recognition on both the original and the degraded versions of all 10,000 images, and statistically compared their results, as summarized in Figure 4d. The statistical results revealed the following recognition outcome categories (Figure S7): (1) Both the original and the degraded images were correctly predicted by the model; (2) The original image was correctly predicted, but the degraded version was misclassified; (3) The original image was misclassified, while the degraded version was correctly predicted; (4) Both the original and the degraded images were incorrectly classified (this category can be further divided into cases where the erroneous labels for the original and degraded images were either identical or different). Among these, cases where the original image was misclassified were attributed to insufficient model accuracy or inherent image ambiguity and were therefore excluded from the assessment of DNA encoding robustness. For cases where the original image was correctly identified, the recognition results for the degraded images were categorized as either correct or incorrect. Here, we use the prediction accuracy (PA), defined as the proportion of cases where the degraded image was correctly predicted given that the original image was correctly classified, to quantitatively assess the impact of image degradation caused by DNA strand loss on recognition performance. We quantified the variation in prediction accuracy across different strand loss rates, as shown in Figure 4e. Although the accuracy gradually decreased as the loss rate increased, it remained above 90% even with a 20% loss. This indicates that image degradation caused by strand loss has minimal impact on recognition performance when using the PJ encoding scheme, and further demonstrates the strong suitability of this algorithmic architecture for DNA-based storage of AI databases.

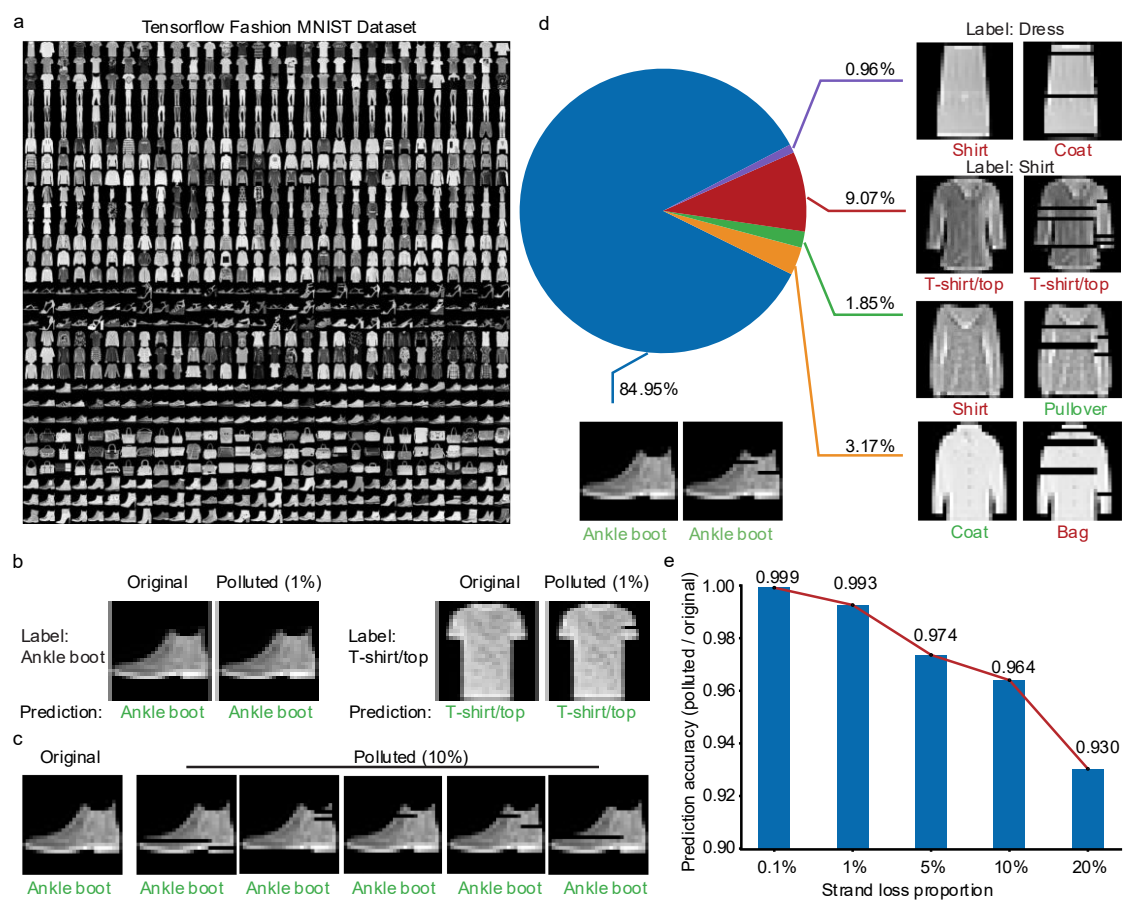

Figure 4 Noise-immune DNA storage for AI pattern recognition. (a) A TensorFlow model trained on the Fashion MNIST dataset[30], which contains 70,000 images of various types of clothing, each with a corresponding category label. (b) Recognition results of the AI model for both the original image and the degraded image (strand loss rate: 1%). (c) Recognition results of the AI model for a boot image subjected to multiple independent rounds of random degradation (strand loss rate: 10%). (d) Statistical outcomes of the AI model's recognition of 10,000 images from the test set after degradation (strand loss rate: 10%). Each category is illustrated with two sample images: the left image represents the original, and the right image represents the degraded version. Text below each image indicates the model's prediction, with green indicating correct predictions and red indicating errors. (e) Change in prediction accuracy (PA) as a function of increasing strand loss rate.

**Experimental validation of PJ code for reliable DNA storage**

Having validated the PJ codec in simulation, we proceeded to experimental demonstration using synthetic DNA strands (Figure 5a). A 285 × 409-pixel cat image (approximately 100 KB of data) was encoded into a DNA strand matrix and a DNA library comprising 5,985 strands was synthesized via chip-based synthesis. This library was then subjected to a complete storage workflow, including PCR amplification and sequencing. The image was successfully decoded and reconstructed from the sequenced data. The recovered image exhibited a high SSIM (0.98) to the original, with minor defects (0.01% erroneous pixels). These errors are likely attributable to imperfections in DNA synthesis, strand loss during storage and PCR amplification, and insufficient sequencing depth. To assess the robustness of the PJ codec against strand loss, we randomly removed approximately 10% of the DNA strands from the encoded library to mimic a strand-loss condition. The depleted library was then synthesized, sequenced and decoded, followed by AI-based inpainting. The raw decoding of this depleted library yielded an image with noticeable artifacts relative to the intact library (Figure 5b). Despite this, the image remained decodable under such high strand loss and its visual quality was markedly improved by inpainting. Next, we tested the resilience of PJ-coded DNA data against extreme environmental disturbance using accelerated aging and high-intensity X-ray irradiation. For accelerated thermal aging, the DNA library was incubated at 95°C for two days. This treatment is thermodynamically equivalent to the strand loss expected from natural degradation over approximately 250 years under ambient conditions[19,31]. Following the treatment, we successfully recovered the degraded image and further enhanced its quality through AI-based inpainting (Figure 5c and Figure S8). We further challenged the library with high-dose X-ray irradiation, which can compromise conventional storage media, such as HDDs, SSDs, flash memory, and magnetic tapes. After exposure to 10 MGy of X-ray for 1 h, the stored data were successfully retrieved (Figure 5d and Figure S9), and the inpainting restored regions affected by damage. Quantitative analysis confirmed that while environmental stressors reduced raw retrieval quality by different levels, the AI-driven approach greatly compensates for degradation-induced information loss (Fig. 5e). This combination of biochemical encoding and computational repair ensures reliable data recovery under harsh environmental stressors, highlighting a path toward fault-tolerant molecular archiving.

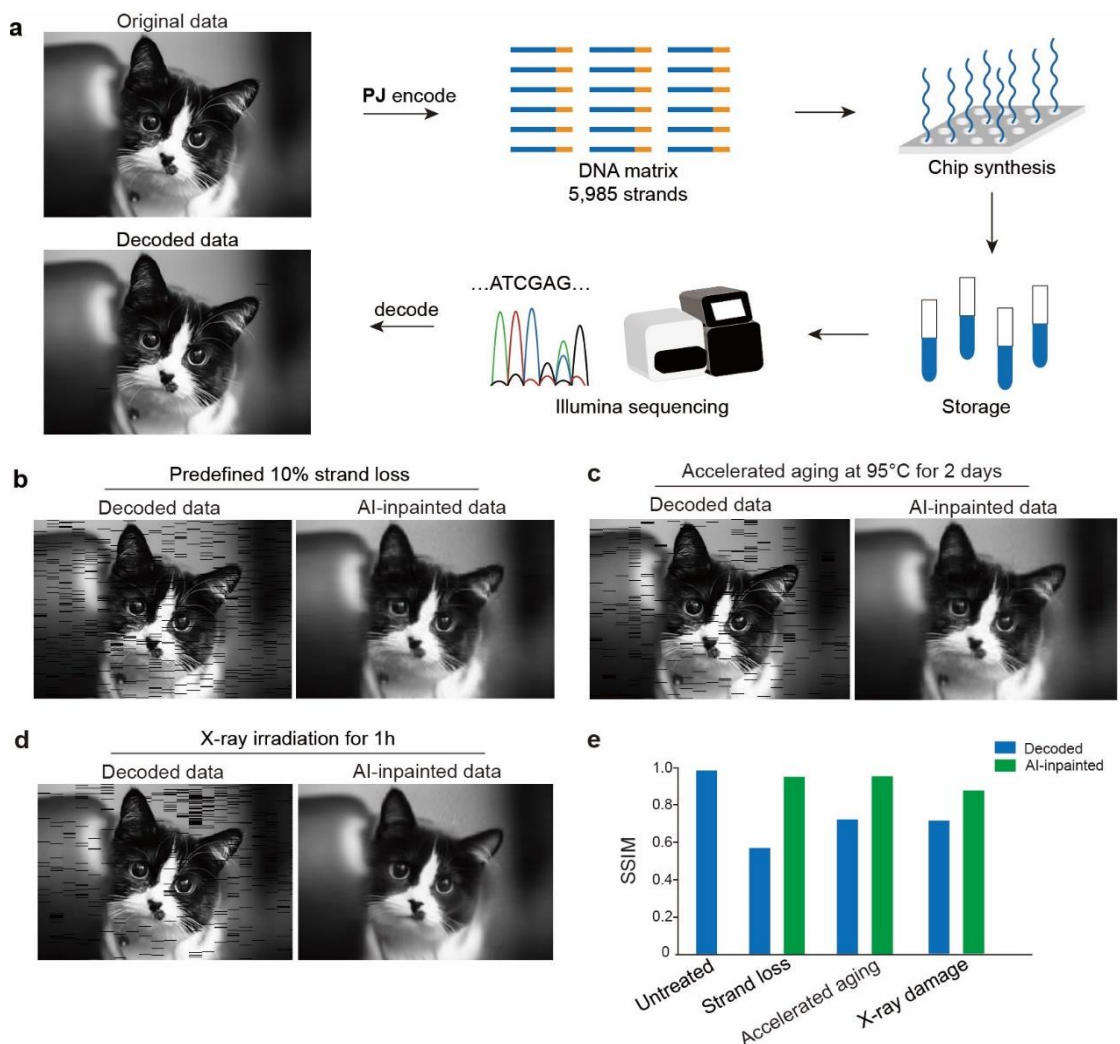

Figure 5 Experimental validation of the PJ codec for noise-immune DNA storage of digital files. (a) Practical workflow of DNA storage using the PJ Code. The image file is successfully recovered after being processed through DNA library synthesis, storage, and next-generation sequencing. (b) Image reconstruction under predefined ~10% strand loss. Decoded image (left) and the corresponding AI-inpainted reconstruction (right). (c) Image restoration after thermal aging at 95°C for two days. Decoded image (left) and the corresponding AI-inpainted reconstruction (right). (d) Image restoration after X-ray irradiation for 1 h. Decoded image (left) and the corresponding AI-inpainted reconstruction (right). (e) SSIM analysis of retrieved images. Comparison of SSIM between raw decoded (blue) and AI-inpainted (green) data for untreated, strand-loss, accelerated-ageing and X-ray–damaged libraries.

**Discussion**

We have presented a noise-immune and AI-compatible encoding algorithm for DNA data storage. The noise immunity ensures robust data preservation independent of specific synthesis and sequencing technologies, while also tolerating material degradation during storage. Information loss in DNA manifests as localized data gaps in the decoded file, which aligns with modern AI inference capabilities for data completion or semantic recognition. This encoding architecture therefore provides a universal interface for digital-molecular information exchange that is decoupled from specific chemical implementations.

The partition-mapping architecture eliminates cross-strand interdependency, thereby enabling unconditional noise immunity and remains robust even under severe strand loss. Conventional DNA encoding strategies typically manage synthesis and sequencing errors by introducing a certain level of physical redundancy or error-correcting codes to ensure information recoverability. This approach is often sufficient for storage under short-term and mild conditions. However, the central value proposition of DNA lies in long-term, high-capacity archiving, where cumulative degradation and episodic exposure to harsh environments can drive error rates and strand loss beyond the designed correction margin, leading to catastrophic file failure. In such cases, the accumulation of errors and strand loss can readily exceed the correction capacity of error-handling codes, ultimately rendering the files stored in DNA irretrievable. By contrast, our noise-immune architecture preserves partial recoverability under arbitrary damage levels, thereby reducing the risk of total data loss and supporting the reliability required for large-scale deployment. In addition, as this approach preserves image recoverability and yields a similarity metric that decreases predictably with increasing loss, it can be developed into a standard assay for evaluating the impact of environmental factors (such as enzyme activity, temperature, pH, and protective coatings)[32] on the preservation longevity of DNA-based information storage.

Within each strand, the tunable Jump-Rotating (JR) rule balances the demands of high information density with the biochemical constraints of synthesis and sequencing technologies. This rule integrates the higher-density quaternary direct encoding into the ternary rotating framework, thereby alleviating the stringent sequence constraints imposed by rotating rules and minimizing the sacrifice in storage density during the encoding process. It exhibits strong compatibility with diverse chemical implementations and offers robust generalization capability for future advancements. For example, setting the jump length to zero makes it compatible with enzymatic synthesis systems[22] that lack precise control over homopolymer extensions. Conversely, if future molecular engineering overcomes limitations on consecutive nucleotide length, setting the jump length to infinity would enable the ultimate encoding density of 2 bits/nt.

The PJ encoding framework reimagines DNA not merely as a high-density hard drive, but as a modern analogue to the Oracle bone script, an enduring substrate capable of carrying civilization's memory through millennia. By rendering data immune to electromagnetic attacks and long-term ageing, localized loss rather than catastrophic file failure, enabling progressive reconstruction instead of all-or-nothing retrieval. This paradigm establishes DNA as a truly immutable medium for the long-term preservation of valuable knowledge, immune to the disturbances that threaten silicon-based storage.

**Acknowledgements**
This work was financially supported by the National Key R&D Program of China (2023YFA0915200), the National Natural Science Foundation of China (T2188102, U24A20497), Shanghai Pilot Program for Basic Research -Shanghai Jiao Tong University (21TQ1400222), Shanghai Municipal Science and Technology Major Project, the New Cornerstone Science Foundation, and the LU JIAXI International team program supported by the K. C. Wong Education Foundation.